\documentclass[prd,twocolumn,showpacs]{revtex4}
\usepackage{graphicx}

\newcommand{\be}{\begin{equation}}
\newcommand{\ee}{\end{equation}}
\newcommand{ \bea}{\begin{eqnarray}}
\newcommand{ \eea}{\end{eqnarray}}
\newcommand{ \mysmall}[1]{\scriptscriptstyle #1} 
\newcommand{ \mz}{M_{\mysmall{Z}}}
\newcommand{ \mw}{M_{\mysmall{W}}}
\newcommand{ \eq}[1]{Eq.~(\ref{eq:#1})}
\newcommand{ \gev}  {\mbox{ GeV}}

\newcommand{ \smallmax}{{\rm\scriptstyle max}}

\newcommand{ \emax}{E_\smallmax}
\def \fr {\frac}

\begin{document}

\title{Observations on the Radiative Corrections to Pion 
\boldmath $\beta$-Decay \unboldmath}


\author{M.~Passera} 
\affiliation{Istituto Nazionale Fisica Nucleare,
Sezione di Padova, 35131, Padova, Italy}
\email{passera@pd.infn.it}

\author{K.~Philippides}
\affiliation{Vocational Lyceum of Rodopolis (EPAL),
62055 Rodopolis, Serres Greece}
\email{philip@physics.auth.gr}

\author{A.~Sirlin}
\affiliation{Department of Physics, New York University, 
10003 New York, NY, USA}
\email{alberto.sirlin@nyu.edu}

\begin{abstract} 
\noindent 
We find that, in the local V--A theory, the radiative corrections to pion $\beta$-decay involving the weak vector current, when evaluated in the current algebra (CA) formulation in which quarks are the fundamental underlying fields, show a small difference with the more elementary calculations based directly on the pion fields. We show that this difference arises from a specific short-distance effect that depends on the algebra satisfied by the weak and electromagnetic currents. On the other hand, we present a simple theoretical argument that concludes that this difference does not occur when the CA formulation is compared with the chiral perturbation theory ($\chi$PT) approach. Comparisons with previous studies, and with a more recent calculation based on $\chi$PT, are included. We also briefly review the important differences between the results in the local V--A theory and the Standard Model.
\end{abstract} 


\maketitle

\section{Introduction}
\label{sec:INTRO}

Pion $\beta$-decay, 
$\pi^+ \rightarrow \pi^0 + e^+ + \nu_e$,
and its charge conjugate, 
$\pi^- \rightarrow \pi^0 + e^- + \bar{\nu_e}$,
are processes of very special interest because, in some sense, they may be regarded as the simplest examples of superallowed $0 \to 0$ Fermi transitions in $\beta$-decay. In particular, their interpretation is devoid of the complications of nuclear structure that affect the traditional superallowed $\beta$-decays of nuclei. On the other hand, their branching ratio, $(1.036 \pm 0.006) \times 10^{-8}$~\cite{PDG}, is extremely small and, as a consequence, the experimental measurement of their decay rate is much less accurate than in the case of their nuclear counterparts.

On the theoretical side, we recall that the superallowed Fermi transitions have played a fundamental role in motivating the conserved vector current hypothesis (CVC)~\cite{Feynman:1958ty,Gershtein:1955fb} and in testing the unitarity of the Cabibbo-Kobayashi-Maskawa (CKM) matrix~\cite{Czarnecki:2004cw,Towner:2007np}.

Radiative corrections to pion $\beta$-decay were first evaluated in the framework of the local V--A theory of weak interactions that preceded the Standard Model (SM), neglecting the effect of the strong interactions and identifying the pions as the basic fields in the interaction Lagrangian~\cite{Terentev,Chang,Arbuzov:1994cw}. An alternative, more general and powerful Current Algebra (CA) approach, that takes into account the effect of the strong interactions, was subsequently developed to study the photonic corrections to $\beta$-decay in the local V--A theory~\cite{Bjorken:1966jh,Abers:1968zz} and the corresponding electroweak corrections in the SM~\cite{Sirlin:1977sv}. It is essentially based on the Ward identities associated with the time-time and time-space components of the CA of weak and electromagnetic currents, which are expressed in terms of the quark fields. Thus, in this approach quarks are identified with the fundamental underlying fields.

An important result of the CA formulation is that, if very small contributions of $O(\alpha E/M)$ are neglected, where $E$ is the charged lepton energy and $M$ a hadronic mass, the photonic corrections to $\beta$-decay involving the hadronic vector current are not affected by the strong interactions~\cite{Abers:1968zz,Sirlin:1977sv}. In particular, one expects that the photonic corrections to pion $\beta$-decay involving the hadronic vector current are essentially the same when evaluated in the CA formulation or using more elementary calculations based directly on the pion fields (PF).

The aim of this paper is to carry out a detailed comparison of the photonic corrections to pion $\beta$-decay arising from the hadronic vector current, as evaluated in the two schemes described above, namely in the PF-based calculation and in the CA formulation. 

In Section II we present our evaluation in the first scheme. In Section III we compare the result with that of the CA formulation, which can be gleaned from Refs.~\cite{Sirlin:1977sv,Sirlin:1967zza}. We find that the two calculations are not identical, but differ by a small, finite term of $O(\alpha)$. We then show that this difference arises precisely from a specific short-distance contribution that depends on both the time-time and space-space components of the CA of the weak and electromagnetic currents. On the other hand, we also present a simple theoretical argument that leads to the conclusion that this difference is no longer present in the comparison between the CA and chiral perturbation theory ($\chi$PT) formulations.

Finally, we briefly discuss the additional important differences that emerge between the local V--A theory and the SM formulation of the corrections. 

\section{Pion Fields' Calculations}	

We consider an effective low energy theory in which pions are regarded as the fundamental fields. In particular, a local V--A theory with the additional CVC hypothesis \cite{Feynman:1958ty,Gershtein:1955fb}. In this framework, the pion $\beta$-decay rate can be cast in the form
\be
	\Gamma = \Gamma_0 \left(1+\delta_\pi \right) \!,
\ee
where $\Gamma_0$ is the leading-order width and  $\delta_\pi$ represents the effect of the radiative corrections. The former is \cite{KallenBook,Sirlin:1977sv}
\be
	\Gamma_0 = \frac{G_F^2 |V_{ud}|^2 \Delta^5}{\pi^3}  \, f(\mu, \Delta)
	\left[1-\fr{\Delta}{2M_{\pi}} \right]^3 \!\!\!,
\ee
where $\mu =m_e/\Delta$, $m_e$ is the electron mass, $M_{\pi}$ and $M_{\pi^0}$ are the masses of the charged  and neutral pions, $\Delta=M_{\pi}-M_{\pi^0}$, 
$G_F=1.1663788(7) \times 10^{-5}\gev^{-2}$~\cite{Webber:2010zf}
is the Fermi coupling constant, and $V_{ud}$ is the CKM matrix element. The function $f(\mu, \Delta)$, up to the leading correction in an expansion in powers of $\Delta^2 / (M_{\pi}+M_{\pi^0})^2 \sim 10^{-4}$, is 
\begin{eqnarray}
	\lefteqn{f(\mu, \Delta) =
	\fr{\sqrt{1-\mu^2}}{30} \left( 1- \fr{9}{2}\mu^2- 4\mu^4 \right) }\nonumber\\
	&& +  \fr{\mu^4}{4} \ln \! \left(\fr{1+\sqrt{1-\mu^2}} {\mu} \right)  
	-\fr{1}{70} \fr{\Delta^2}{(M_{\pi}+M_{\pi^0})^2}.
\label{eq:fmu}           
\end{eqnarray}
%

\begin{figure}
\includegraphics[width=85mm,angle=0]{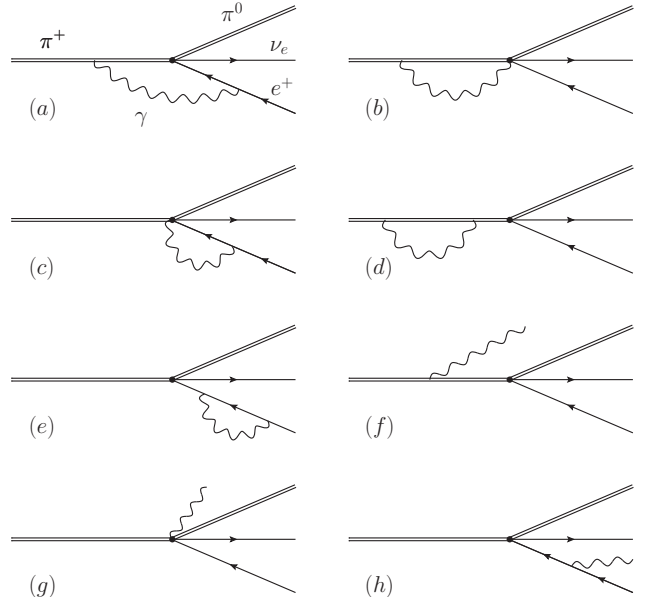}
\vspace{0cm}\caption{The diagrams contributing to the $O(\alpha)$ corrections to the pion $\beta$-decay width in the local V--A theory: $(a)$--$(c)$, vertex corrections; $(d)$,$(e)$, wave-function renormalizations; $(f)$--$(h)$, bremsstrahlung graphs.}
\label{fig:diagrams}
\end{figure}

The $O(\alpha)$ radiative corrections to the pion $\beta$-decay width in the local V--A theory are given by the sum
\be
	\delta_\pi = \fr{\alpha}{\pi} \left( \delta^V +  \delta^{S_\pi} + \delta^{S_e}
	+ \delta^W + \delta^{brem} \right) \!,
\ee
where $\delta^V,$ $\delta^{S_\pi}$ and $\delta^{S_e}$ are the vertex corrections of Figs.~1~$(a)$,$(b)$,$(c)$, respectively, and $\delta^W$ denotes the contributions of the charged pion and charged lepton wave-function renormalizations (Figs.~1~$(d)$,$(e)$). The term $\delta^{brem}$ is the contribution of inner bremsstrahlung 
(Figs.~1~$(f)$,$(g)$,$(h)$).
If in these corrections we neglect terms of 
	$O[(\alpha/\pi) \, \Delta / M_{\pi}]$ and
	$O[(\alpha/\pi) \, m_e/M_\pi]$, 
we obtain the following explicit expressions:
\begin{eqnarray}
\delta^V &=& \fr{1}{I(\epsilon)} \int_{\epsilon}^{1} dx \,
	  \beta \, x^2 \, (1-x)^2 \, \biggl\{
	\fr{7}{4} \ln \fr{\Lambda}
	{M_\pi}  + 
\nonumber \\ && 
	\!\!\!\!\!\!\!\!\!\!\! + \fr{17}{16} + \ln \fr{M_\pi}{m_e} + \frac{1}{\beta} \,
	\mbox{Li}_2 \left(\fr{2\beta}{\beta-1}
	\right) +  \fr{1}{2 \beta} \times
\nonumber \\ && 
	\!\!\!\!\!\!\!\!\!\!\!  \times \left[ \beta^2 \! 
	+ \fr{1}{2}\ln \! \left( \frac{1+\beta}{1-\beta} 
	\right) \! + \ln \fr{\lambda^2}{m_e^2} \right]
	\! \ln \! \left(\frac{1+\beta}{1-\beta} \right) \!\biggr\}, \\
\delta^{S_\pi} &=&  \fr{3}{4} \ln \fr{M_\pi}{\Lambda} - 
	\fr{9}{16},     \\
\delta^W &=& 2 \ln \fr{m_e}{\lambda} - \fr{1}{2} \ln \fr{m_e}{\Lambda} -\fr{3}{2},
\end{eqnarray}
where $\beta=\sqrt{1- \epsilon^2/x^2}$, $\epsilon=m_e/\emax$, $x=E/\emax$, $E$ is the positron energy, ranging from $m_e$ to the end-point energy
\bea
	\emax = \Delta \left[ 1- \fr{\Delta}{2M_{\pi}} \left( 1 -\mu^2 \right) \right] \!, 
\\
	I(\epsilon) = \!\int_{\epsilon}^{1} dx \, \beta \, x^2 \, (1-x)^2  \,=\,  f(\epsilon,0),
\eea
and $\mbox{Li}_2(z) = - \int_0^z (dt/t) \ln(1-t)$ is the dilogarithm. Since $\Delta/M_\pi = 0.0329$ is quite small, we note that
$\emax \approx \Delta$, 
$\epsilon \approx \mu$, and 
$I(\epsilon) \approx f(\mu,\Delta)$.
The parameter $\lambda$ is the infrared regulator of the photon and $\Lambda$ is the Feynman ultraviolet (UV) cutoff. We also note that the correction $\delta^{S_e}$ is proportional to $m_e/M_{\pi}$ and was therefore neglected.
For the emission of one real photon, integrated over its entire energy range, we obtain
\begin{eqnarray}
\lefteqn{ \delta^{brem} =
	\fr{1}{I(\epsilon)} \int_{\epsilon}^{1} \! dx \, \beta  \, x^2 \, 
	(1-x)^2 \biggl\{4 -\fr{2-2x}{3x} \, +} 
\nonumber \\ && 
	-2 \ln (2-2x) - 2\ln \fr{\emax}{\lambda} - \fr{1}{\beta} \, \mbox{Li}_2 \! \left( 
	\fr{2\beta}{1+\beta} \right) \!+ 
\nonumber \\  && 
	+ \, \fr{1}{\beta} \biggr[ \ln \fr{\emax}{\lambda} + \ln \left(2-2x \right) 
	-\fr{1}{4} \ln \! \left( \fr{1+\beta}{1-\beta} \right) \!+ \nonumber \\ 
	& & + \, \fr{(1-x)(1+7x)}{24 x^2}   -1   \biggr]	
	\ln  \left( \fr{1+\beta}{1-\beta} \right) \biggr\}.
\end{eqnarray}

The total effect of the $O(\alpha)$ radiative corrections in the local V--A theory can therefore be cast in the form
\begin{eqnarray}
\label{eq:deltapi}
\lefteqn{ \delta_\pi = \frac{\alpha}{\pi} 
	\frac{1}{I(\epsilon)} \int_{\epsilon}^{1} dx \, \beta \, x^2 \,(1-x)^2
	\biggl\{ 3 - 2 \ln (2-2x) \, +}
\nonumber \\   &&   
	+ \, \frac{3}{2} \ln 
	\frac{\Lambda}{\emax}
	+ \frac{1}{2} \ln \epsilon - \frac{2-2x}{3x} 
	- \fr{2}{\beta} \mbox{Li}_2 \! \left( \frac{2\beta}{1+\beta} \right)+ 
\nonumber \\   && 
	+ \, \fr{1}{\beta}  \, \ln  \left( \frac{1+\beta}{1-\beta} \right) 
	\biggl[ \frac{\beta^2}{2} + \ln \left(\frac{2}{1+\beta}\right) + 
\nonumber \\  & & 	
	+ \, \ln \left( \frac{1-x}{x} \right) 
	+ \fr{(1-x)(1+7x)}{24 x^2} -1
	\biggr] \biggr\}. 
\end{eqnarray}
This result is infrared finite and, in accordance with the Kinoshita--Lee--Nauenberg (KLN) theorem~\cite{Kinoshita:1958ru,KLN}, free of mass singularities; indeed, after explicit integration, the sum of all $\ln\epsilon$ terms in Eq.~(\ref{eq:deltapi}) cancels in the limit $\epsilon \rightarrow 0$ and we obtain 
\be
\label{eq:deltapilimit}
\lim_{\epsilon \to 0}  \delta_{\pi} =  \frac{\alpha}{\pi} \left[ \frac{3}{2}\ln \left(\frac{\Lambda}{2 \emax} \right) - \frac{2\pi^2}{3} +\frac{177}{40} \right].
\ee
On the other hand, \eq{deltapi} is logarithmically divergent in the UV domain, a well known property of the radiative corrections to $\beta$-decay in the local V--A theory~\cite{Kinoshita:1958ru}.  It is also interesting to note that, to leading order, $\delta_\pi$ is independent of the pion mass.

The leading-order formula for $\delta_{\pi}$ has already been published in (at least) three articles~\cite{Terentev,Chang,Arbuzov:1994cw}. Alas, these formulae do not agree (although some of the provided numerical values do -- see below). Our result in \eq{deltapi} agrees with the Erratum of Ref.~\cite{Arbuzov:1994cw}, except for small terms of $O[(\alpha/\pi) \, \epsilon^2]$ and $O[(\alpha/\pi) \, \Delta / M_{\pi}]$. Equation (18) of Ref.~\cite{Terentev}, which provides the full expression for the radiative corrections to the positron spectrum, appears to contain a typographical error, but its integration in the $\epsilon \to 0$ limit agrees with our \eq{deltapilimit}. Also Eq.~(15) of Ref.~\cite{Chang} for $-\delta_{\pi}$ contains typographical errors. It diverges as $m_e \to 0$ and, therefore, does not satisfy the theorem on the cancellation of mass singularities in integrated observables~\cite{Kinoshita:1958ru,KLN}. Thus, it differs from our Eqs.~(\ref{eq:deltapi},\ref{eq:deltapilimit}), which satisfy this important theorem. 

The result of the numerical integration of \eq{deltapi} is 
	$\delta_{\pi} = (1.14, 2.73)\%$,
where the first value in parentheses is obtained with $\Lambda = m_p$, the proton mass, and the second one with $\Lambda = \mz$, the $Z$ boson mass. The corresponding values obtained from Eq.~(16) of the Erratum of Ref.~\cite{Arbuzov:1994cw} are
	$\delta_{\pi} = (1.20, 2.79)\%$.
They are in agreement with ours, apart from small contributions of $O[(\alpha/\pi) \, \epsilon^2]$ and $O[(\alpha/\pi) \, \Delta / M_{\pi}]$. The $m_e \to 0$ limit (cf.\  our \eq{deltapilimit} and Ref.~\cite{Terentev}) leads to (1.12, 2.71)\%, a result very close to that obtained from the complete \eq{deltapi}.

It is also interesting to note that, more recently, the radiative corrections to pion $\beta$-decay have been studied in the framework of $\chi$PT~\cite{Cirigliano:2002ng}. In order to compare numerically the final result in that paper with ours, in Eq.~(5.4) of Ref.~\cite{Cirigliano:2002ng} we replace the short-distance electroweak enhancement factor according to 
$S_{\rm EW}(M_{\rho},\mz)  \to 1 + (3\alpha/2\pi) \ln(\Lambda/M_{\rho})$.
In analogy with our \eq{deltapi}, this corresponds to using the leading, $O(\alpha)$ approximation for $S_{\rm EW}$, keeping only the contribution from the vector current to this factor, and replacing $\mz \to \Lambda$. Using this change, and taking into account the additional results in that paper, we find that Eq.~(5.4) of Ref.~\cite{Cirigliano:2002ng} leads to $\delta_{\pi} = (1.10 \pm 0.10, 2.71 \pm 0.10)\%$, which is also in good agreement with the values derived from our \eq{deltapi}.

\section{Comparison with the current algebra formulation}

In the SM, the electroweak corrections to the superallowed Fermi transitions, evaluated in the CA formulation, are given in Eq.~(7.7) of Ref.~\cite{Sirlin:1977sv}. In particular, the corrections arising from the hadronic vector current involve the first two terms in that expression, namely 
$P^{0} d^{3}p_e [3 \ln (\mz/m_p) + g(E,E_m)]$, 
where $P^{0}$ is the uncorrected decay probability, $E$ and $p_e$ are the energy and three-momentum of the charged lepton, $E_m$ the end-point energy, and the function $g(E,E_m)$, defined in Eq.~(20b) of Ref.~\cite{Sirlin:1967zza}, describes the corrections to the positron or electron spectrum in $\beta$-decay. This result has three important properties: i) it is UV convergent, which reflects the fact that the SM is a renormalizable theory, ii) it leads to a large correction of $O(4\%)$ to the decay rate, iii) the proton mass $m_p$ cancels and one finds out that the complete expression $3\ln(\mz/m_p) + g(E,E_m)$ contains no reference to the hadronic sector. We also recall that, in the CA formulation, the $O(\alpha)$ SM corrections to the Fermi amplitude can be obtained from those in the local V--A theory by the simple substitution $\Lambda \to \mz$~\cite{Sirlin:1977sv,Sirlin:1974ni}. This implies that, in the CA formulation, the radiative corrections arising from the hadronic vector current in the local V--A theory are given by
\be
(\delta P)_{CA} = P^{0} d^{3}p_e (\alpha/2\pi) [3 \ln(\Lambda/m_p) + g(E,E_m)].
\label{eq:deltaPCA}
\ee
In order to compare \eq{deltaPCA} with \eq{deltapi}, we insert the explicit expression for $g(E,E_m)$, and integrate over the charged lepton momentum. The comparison shows that these two lengthy and complicated expressions are indeed very close, but not identical. In fact, we find that the corrections evaluated in the CA formulation lead to an additional contribution $-3 \alpha/8\pi$ to $\delta_{\pi}$, not present in the PF calculation given in \eq{deltapi}.

We now show that this difference can be traced to a specific short-distance contribution to the radiative corrections, namely the amplitude
\be
Y = i \, \frac{G_F}{\sqrt{2}} \, \frac{\alpha}{2\pi^3} \,\, L_{\mu} \int \! d^{4} k 
\, \frac{\Lambda^2}{\left(\Lambda^2 - k^2 \right)^2}
\frac{k^{\mu}}{k^2} \,\, T^{\lambda}_{\lambda}(k),
\label{eq:Y}
\ee
where $L_{\mu}$ is the leptonic current and
\be
T^{\lambda \rho}(k)  = \int \! d^{4}x \,\,  e^{ik\cdot x} \langle p' | T[J_{\gamma}^{\lambda}(x) V_W^{\rho}(0)] | p \rangle
\ee
is the Fourier transform of the time-ordered product of the hadronic electromagnetic and vector currents $J_{\gamma}^{\lambda}$ and $V_W^{\rho}$ (see, for example, Eq.~(C 25) of Ref.~\cite{Sirlin:1977sv}, where $\mw^2$ plays the role of the cutoff $\Lambda^2$). The fact that $Y$ is model dependent was already pointed out in Ref.~\cite{Abers:1968zz}.

As $\Lambda^2 \to \infty$, the only non-vanishing terms in $Y$ arise from the contributions to $T^{\lambda}_{\lambda}(k)$ of $O(1/k)$ in the large $k$ limit. A simple way to find these contributions is to employ the expression for the Bjorken-Johnson-Low limit, namely the limit for large $k^{0}$ for fixed $\vec{k}$~\cite{Bjorken:1966jh,Johnson:1966se}. It's given by
\be
T^{\lambda}_{\lambda}(k) \to  \frac{i}{k^{0}} \int \! d^{4}x \,\, e^{ik \cdot x} \delta \left(x_0 \right)
\langle p' | [J_{\gamma}^{\lambda}(x),V_{W \lambda}(0)] | p \rangle.
\label{eq:Tll1}
\ee
In the CA formulation, $J_{\gamma}^{\lambda} = \bar{\psi} \gamma^{\lambda} Q \psi$, $V_W^{\lambda} = \bar{\psi} \gamma^{\lambda} C_{-} \psi$, where $\psi$ is a column vector whose components are the quark fields $(u,c,t,d,s,b)$ and $Q$ is the electric charge matrix. In the three-generation case, $C_-$ is a $6 \times 6$~matrix of the form
\be
	C_{-} = \left( \begin{array}{c|c} 0 & 0 \\ \hline U^{\dagger} & 0 \end{array} \right) \!,
\ee
where $U$ is the $3 \times 3$~unitary CKM matrix (with this definition, $V_W^{\lambda}$ is the $\Delta Q = -1$ hadronic vector current, in analogy with Ref.~\cite{Sirlin:1977sv}). Evaluating the equal-time commutator, one finds
\be
\delta(x_0) [J_{\gamma}^{\lambda}(x),V_{W \lambda}(0)] = 2 \delta^{(4)} \! (x) V_W^{0}(0),
\label{eq:commutator}\ee
\be
T^{\lambda}_{\lambda}(k) \to  \frac{2i}{k^{0}}  \langle p' | V_W^{0}(0) | p \rangle,
\ee
or, in covariant form:
\be
T^{\lambda}_{\lambda}(k) \to  \frac{2i}{k^2}  \langle p' | k^{\lambda} V_{W \lambda}(0) | p \rangle.
\label{eq:Tll2}
\ee
Inserting \eq{Tll2} in \eq{Y} and performing the integration, we find
\be
Y_{CA} = -\left( \frac{\alpha}{8\pi}\right) M^{0},
\label{eq:YCA}
\ee
where $M^{0}$ is the contribution of the hadronic vector current to the zeroth-order amplitude for the $\beta$-decay under consideration. This gives a contribution 
$-(\alpha/4\pi) \Gamma^{0}$ 
to the decay rate. Instead, in the PF case we have
\bea
J_{\gamma}^{\lambda}(x) &=& i \left[ \phi^\dag \partial^{\lambda}\phi - \partial^{\lambda} \phi^\dag \phi \right] \! (x), 
\label{eq:EMC}
\\
V_W^{\lambda}(x) &=& i \sqrt{2} \left[ \phi^{0} \partial^{\lambda}\phi - \partial^{\lambda} \phi^{0} \phi \right] \! (x),
\label{eq:HadV}
\eea
where $\phi(x)$ and $\phi^{0}(x)$ are the $\pi^{+}$ and $\pi^{0}$ fields. Inserting Eqs.~(\ref{eq:EMC}, \ref{eq:HadV}) in \eq{Tll1} and evaluating the equal-time commutator, we obtain
\be
	\delta(x_0) [ J_{\gamma}^{\lambda}(x), V_{W \lambda}(0)] = - \delta^{(4)}\! (x) V_W^{0}(0),
\ee
which is $-(1/2)$ the value found in \eq{commutator}. Correspondingly, in the PF case, we have
\be
	Y_{PF} = \left( \frac{\alpha}{16\pi} \right) M^{0}.
\label{eq:YPF}
\ee
Combining Eqs.~(\ref{eq:YCA},\ref{eq:YPF}):
\be
Y_{CA} - Y_{PF} = - \left( \frac{3 \alpha}{16\pi} \right) M^{0}.
\ee
This leads to a change $-(3\alpha/8\pi)\Gamma^{0}$ in the corrections to the decay rate and explains the difference of $-3\alpha/8\pi$ we encountered between the CA and PF calculations of $\delta_{\pi}$! It is important to note that this effect arises because the asymptotic behaviour of $T^{\lambda}_{\lambda}(k)$ for large $k$, as shown in \eq{Tll1}, is governed by both the time-time and space-space components of the current algebra of the weak and electromagnetic currents. Indeed, the space-space components of the algebra are different in the CA and PF formulations. On the other hand, this shift leads to a change of only 0.09\% in the theoretical calculation of the decay rate, which is about seven times smaller than the current experimental accuracy of 0.6\%.

Subtracting $3\alpha/8\pi$ from \eq{deltapi}, we see that our CA result is $\delta_{\pi} = (1.05, 2.64)\%$. Thus, within its 0.10\% theoretical error, the $\chi$PT result of Ref.~\cite{Cirigliano:2002ng} agrees with both our CA and PF calculations.

It is important to point out that the small discrepancy we have uncovered between our CA and PF calculations is no longer present in a comparison between the CA and $\chi$PT approaches. The reason is that the amplitude $Y$ that leads to the discrepancy is a short-distance effect (cf.\ \eq{Y} and the discussion following that equation), and in such domain the relevant weak and electromagnetic currents are those of the underlying theory of $\chi$PT, namely the SM. Such currents involve quark rather than pion fields and, consequently, the value of the $Y$ amplitude is expected to be the same in the $\chi$PT and CA formulations. 
We also note that the 0.10\% error in the $\chi$PT calculation of Ref.~\cite{Cirigliano:2002ng} is of the same magnitude as the 0.09\% discrepancy, so at present it is not possible to verify our conclusion by a numerical comparison of the CA and $\chi$PT results.

The electroweak corrections to $\beta$-decay evaluated in the SM present very important differences with the results of the local V--A theory in general, and the PF calculation of $\pi^{+} \to \pi^{0} + e^{+} + \nu_e$ in particular:
i) Since the SM is a renormalizable theory, the corrections are UV convergent and well defined, with $\Lambda = \mz$. The large cutoff leads to sizable corrections of O(4\%) that are phenomenologically necessary to satisfy CKM unitarity. Indeed, the most precise test of this fundamental property is currently satisfied at the 0.06\% level~\cite{Towner:2007np}.
ii) With respect to the PF formulation of the corrections to pion $\beta$-decay, there is another important difference. In the SM, the hadronic axial vector current, $\bar{\psi} \gamma^{\rho} \gamma_5 C_{-} \psi$, does not contribute to the Fermi amplitude at the tree-level but, at the one-loop level, it gives rise to a very significant correction of $O[(\alpha/2\pi) \ln(\mz/M)] \sim 0.5\%$ ($M$ is a hadronic mass of $O(1\gev)$). In fact, this correction plays an important role in precision calculations of nuclear Fermi transitions~\cite{Czarnecki:2004cw,Towner:2007np}. As shown, for example, in Section IV.B of Ref.~\cite{Sirlin:1977sv}, it involves the terms of $O(1/k)$ in the large $k$ expansion of 
$A^{\lambda \rho}(k) = \int d^{4} x \, e^{ik \cdot x} \langle p' | T [J_{\gamma}^{\lambda}(x) A_W^{\rho}(0) | p \rangle$, 
where $A_W^{\rho}(0)$ is the hadronic axial vector current. On the other hand, in the PF formulation, the natural form for the axial vector current is $A_W^{\rho}(x)  = f_{\pi} \partial^{\rho} \phi(x)$, where $f_{\pi}$ is the $\pi_{l2}$ decay constant. Using this expression, the electromagnetic current given in \eq{EMC}, and employing again the Bjorken-Johnson-Low limit, one readily finds that the terms of $O(1/k)$ in $A^{\lambda \rho}(k)$ don't contribute to pion $\beta$-decay. Thus, in the PF approach, it is not possible to generate the large corrections of $O(\ln \mz/M)$ or $O(\ln \Lambda/M)$ to the Fermi amplitude that arise from the axial vector current in the CA formulation.

\section{Summary}

In conclusion, we have studied the radiative corrections to pion $\beta$-decay in the local V--A theory in a framework in which pions are treated as the fundamental fields, and compared the result for the decay rate with the corrections involving the hadronic vector current in the modern Current Algebra formulation, in which quarks are identified as the fundamental underlying fields. We found a small finite difference that we traced to a specific short-distance contribution that involves both the time-time and space-space components of the algebra satisfied by the hadronic electromagnetic and weak currents. On the other hand, we also presented a theoretical argument that concludes that this difference is no longer present in the comparison between the CA and $\chi$PT approaches. Comparisons with previous studies, and with a more recent calculation based on $\chi$PT, were also included. Finally, we briefly discussed the important differences between the electroweak corrections evaluated in the SM with the corresponding calculations in the local V--A theory in general, and the pion fields' treatment of $\pi^{+} \to \pi^{0} + e^{+} + \nu_e$ in particular.

\begin{acknowledgments}

We would like to thank B.~Ananthanarayan, V.~Cirigliano and G.~Degrassi for very useful comments. 
M.P.\  acknowledges the support of the Centre for High Energy Physics at the Indian Institute of Science, Bangalore, during a visit when this work was finalized, the Department of Physics of the University of Padova, the Fondazione Cariparo Excellence Grant {\it LHC and Cosmology}, and the European Programme {\small PITN-GA-2009-237920}, {\it Unification in the LHC Era}. 
The work of A.S.\ was supported in part by the U.S.\ NSF grant {\small PHY-0758032}. 
Feynman diagrams were drawn with JaxoDraw~\cite{Jax}.

\end{acknowledgments}


\end{document}